%% file: _main.tex
\documentclass[conference]{IEEEtran}
\IEEEoverridecommandlockouts

\usepackage{cite}
\usepackage{amsmath,amssymb,amsfonts}
\usepackage{algorithmic}
\usepackage{graphicx}
\usepackage{textcomp}
\usepackage{xcolor}

\usepackage{url}
\usepackage{multirow}
\usepackage{cleveref}

\def\BibTeX{{\rm B\kern-.05em{\sc i\kern-.025em b}\kern-.08em
    T\kern-.1667em\lower.7ex\hbox{E}\kern-.125emX}}
\begin{document}

\title{Lead Instrument Detection from Multitrack Music}

\author{\IEEEauthorblockN{Longshen Ou}
\IEEEauthorblockA{\textit{School of Computing} \\
\textit{National University of Singapore}\\
Singapore \\
oulongshen@u.nus.edu}
\and
\IEEEauthorblockN{Yu Takahashi}
\IEEEauthorblockA{\textit{Research and Development Division} \\
\textit{Yamaha Corporation}\\
Hamamatsu, Japan \\
yu.takahashi@music.yamaha.com}
\and
\IEEEauthorblockN{Ye Wang}
\IEEEauthorblockA{\textit{School of Computing} \\
\textit{National University of Singapore}\\
Singapore \\
wangye@comp.nus.edu.sg}

}

\maketitle

\input{sections/0_abstract}

\input{sections/1_intro}

\input{sections/2_method}
\input{sections/3_experiment}

\input{sections/4_results}

\input{sections/5_conclusion}

\bibliographystyle{IEEEtran}
\bibliography{IEEEabrv,_mybib}

\end{document}

%% file: sections/0_abstract.tex
\begin{abstract}
Prior approaches to lead instrument detection primarily analyze mixture audio, limited to coarse classifications and lacking generalization ability. This paper presents a novel approach to lead instrument detection in multitrack music audio by crafting expertly annotated datasets and designing a novel framework that integrates a self-supervised learning model with a track-wise, frame-level attention-based classifier. This attention mechanism dynamically extracts and aggregates track-specific features based on their auditory importance, enabling precise detection across varied instrument types and combinations. Enhanced by track classification and permutation augmentation, our model substantially outperforms existing SVM and CRNN models, showing robustness on unseen instruments and out-of-domain testing. We believe our exploration provides valuable insights for future research on audio content analysis in multitrack music settings.
\end{abstract}


%% file: sections/1_intro.tex
\section{Introduction}

Audio content analysis is a crucial area of study within audio processing and music information retrieval, focusing on tasks like identification, transcription, and segmentation \cite{lu2002content}. Beyond these foundational tasks, a critical aspect of human musical perception involves identifying the \textbf{lead instrument}---the instrument that captures the listener's attention with its dominant auditory presence. This could range from the lead vocals \cite{burgel2021listening} and guitar solos \cite{goertzel1991rock} in rock and pop, to lead saxophone or trumpet \cite{abesser2014dynamics} as well as drum solos \cite{king2014brief} in jazz. Automating lead instrument detection can not only facilitate the creation of audio thumbnails and music structural analysis but also potentially simplify audio mixing workflows and enhance music recommendation systems.

However, prior research on lead instrument detection has primarily focused on analyzing mixture audio \cite{wieczorkowska2008training, wieczorkowska2010identification, peterschmitt2001pitch, smit2007solo, fuhrmann2009detecting, mauch2011timbre} or isolated single instrument tracks \cite{pati2017dataset, pachet2013reflexive, foulon2014automatic}, limiting their ability to capture high-level, instrument-specific properties like roles and interactions within a song, essential for identifying lead instruments. These studies often involve coarse-level classification, confined to predefined categories like vocals or guitar solos \cite{mauch2011timbre, pati2017dataset}, restricting applicability in real-world settings where any instrument can serve the lead role. Additionally, many works rely on Support Vector Machine (SVM) models \cite{wieczorkowska2008training, wieczorkowska2010identification, peterschmitt2001pitch, smit2007solo, fuhrmann2009detecting, mauch2011timbre, bosch2012comparison, pati2017dataset}, constrained by the necessity to include all potential lead instruments in training data, thus failing to identify new instruments absent from training.


The analysis of multitrack music, where each track contains specific instrument audio in a time-synchronized, multi-stream format, remains significantly underexplored. To our knowledge, there is no existing work that outlines neural network designs specifically for handling multitrack audio inputs in content analysis tasks. While prior studies on automatic mixing tasks have utilized deep learning models \cite{martinez2021deep, steinmetz2021automatic, martinez2022automatic, koo2023music}, these approaches often assume a fixed number and type of instruments, which limits their applicability in real-world scenarios, where track counts and instrument types can vary significantly. Recently, \cite{vanka2024diff} introduced a framework capable of handling arbitrary track combinations for automatic mixing, sharing some design similarities with our model.

The pre-train and fine-tune paradigm with self-supervised learning (SSL) models like wav2vec 2.0 \cite{baevski2020wav2vec} and HuBERT \cite{hsu2021hubert} has revolutionized audio and speech domains, advancing music information retrieval \cite{gu2024automatic} and reducing reliance on in-domain data \cite{gao2023self, ou2022transfer}. Music-specific SSL models have also achieved state-of-the-art results in various tasks \cite{yizhi2023mert}, but their use remains largely limited to single-stream audio. Extending these models to multitrack audio requires adapting single-track architectures to multitrack context and enabling effective cross-track feature integration. Addressing these issues is crucial for enhancing SSL applications in complex multitrack environments.

This paper develops neural network models for detecting the lead instrument at any moment in multitrack music audio, extending conventional vocal and guitar solo detection to any type of lead instrument while adapting to the multitrack setting without assumptions about track count or type. We created two expertly annotated datasets and compared various model designs. Our model uses a shared SSL audio encoder across tracks, with a novel track-wise attention mechanism that aggregates features from each instrument track based on its importance relative to the mixture track. To further enhance performance, we introduced a track permutation augmentation strategy to diversify the training data. Our key contributions include:

\begin{itemize}
    \item \textbf{Initiate the task of lead instrument detection from multitrack music}, with expertly annotated datasets, and strong baseline models across multiple settings, including analyses for both segment-level and frame-level, both multitrack and single mixture tracks.\footnote{\label{sharedfootnote}Please refer to \url{https://github.com/Sonata165/LeadInstrumentDetection} for code, dataset annotations, and model checkpoints.}
    \item \textbf{Superior performance compared to existing models}, including SVM- and CRNN-based approaches, and also generalizable to unseen instruments and new domains.
    \item \textbf{Evaluated multiple model designs}, demonstrating the advantages of the proposed track-wise attention and track classification.
\end{itemize}

%% file: sections/2_method.tex
\section{Method}

\subsection{Problem Definition}
The task involves identifying the lead instrument at any given timestep within a multitrack music recording, composed of time-synchronized audio tracks from different instruments of the same song. We assume only one lead instrument at each timestep, indicated by track number or instrument name. We also assume access to track-wise metadata including instrument type and track ID, as well as a human-produced mixture track, which we utilize to aid the detection.

\subsection{Crafting Datasets}


To establish a foundation for this new task, we created two datasets with expert annotations from two different audio sources. They include an internal dataset named MJN includes multitrack recordings from five live events, each featuring performances by 4 to 6 bands, and the MedleyDB dataset \cite{bittner2014medleydb}, known for its use in various MIR tasks. Annotations were performed by experts using Adobe Audition, who marked lead instruments based on audio playback and observing waveforms. The process involved identifying onsets, offsets, and instrument types, with specific rules for overlapping leads and minimum segment durations. For more details, please refer to our dataset documentation\footref{sharedfootnote}. The resulting datasets were 7.10 h and 5.57 h in duration, with further elaboration on splitting strategies in Section \ref{sec:implementation}. 


We highlight some important properties of the datasets. In both MJN and MedleyDB, frequencies of lead instrument type exhibit a long-tailed distribution, with vocals being the most frequent lead instrument, followed by electric guitar. Notably, the MJN dataset, commonly used open microphones on stage, contains considerable amount of bleeding sound. While MedleyDB offers a greater variety of instrument types---29 in total, with 26 as leads (14 and 13 in MJN), MJN shows a more balanced distribution among lead instruments, where the frequency of the second most common lead is 53\% that of the most frequent, versus 37.1\% in MedleyDB. Additionally, MJN features more frequent lead instrument switches (avg. 23.05 s per change) than MedleyDB (avg. 29.33 s per change).

\begin{figure}[tb]
\centerline{\includegraphics[width=\columnwidth]{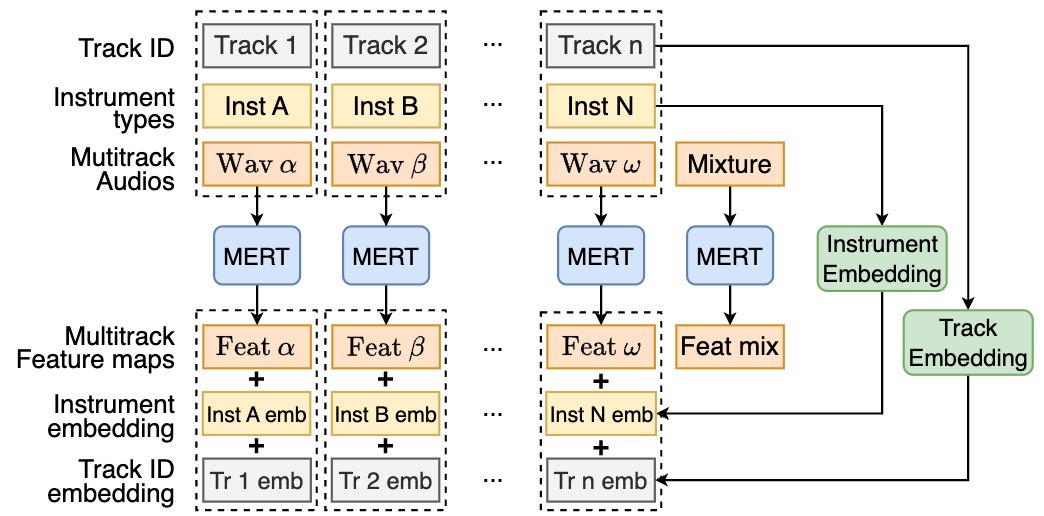}}
\caption{Encoding information for each track.}
\label{fig:audio_enc}
\end{figure}

\begin{figure}[tb]
\centerline{\includegraphics[width=0.85\columnwidth]{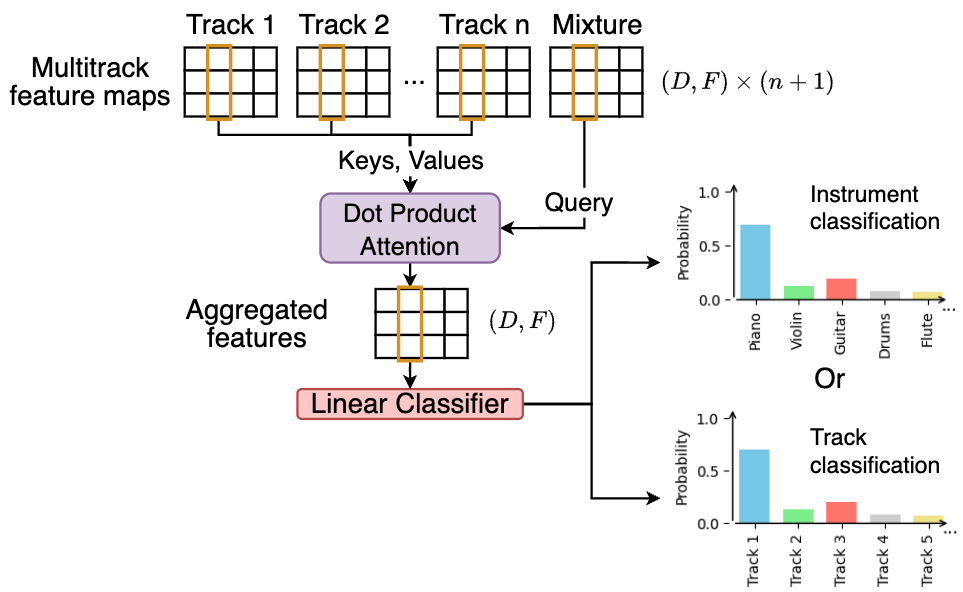}}
\caption{Track-wise frame-level attention and subsequent classification.}
\label{fig:track_attn}
\end{figure}

\subsection{Lead Instrument Detection Model}

Our model, depicted in Figures~\ref{fig:audio_enc} and \ref{fig:track_attn}, integrates an audio encoder with an attention-based classifier. Audio from each track is processed through a shared encoder to generate track-wise feature maps. We employ MERT \cite{yizhi2023mert}, a music-specific self-supervised learning (SSL) model, as our audio encoder due to its robust performance in timbre and pitch-related tasks. Instrument types and track IDs are embedded and added to the feature maps: instrument embeddings reduce the encoder's burden by delegating timbre-to-instrument mapping, while track embeddings enables track classification. A track-wise attention mechanism aggregates these features into a single feature map, which a frame-wise linear classifier uses to produce the final classification results.



To aggregate multitrack information effectively, we designed a frame-level track-wise dot-product attention mechanism. For each frame, the mixture track serves as the query, while individual instrument tracks act as keys and values. This setup enables the dot product to perform a nuanced comparison of each instrument track against the global mixture, determining the importance level of each track at every timestep. Subsequently, features from all tracks are aggregated into a single feature map through a weighted sum, with weights derived from the initial comparisons’ attention scores. This aggregation prioritizes tracks that contribute most significantly to the overall auditory effect. This design emphasizes track-to-mixture content comparisons to highlight the prominence (or relevance) of specific sound sources within a complex multitrack environment, mirroring the selective auditory attention of humans.

Direct instrument classification faces challenges with generalizability. It fails to accurately classify untrained instrument types, instead misclassifying them as similar-sounding instruments. Moreover, it cannot distinguish between multiple instances of the same instrument type across different tracks, which limits its practical applicability. To overcome these issues, we adjusted the classification scheme from instrument types to track IDs of lead instruments. This modification not only enable generalization to unseen instruments but also improves performance in out-of-domain testing.


Despite the potential for higher generalization, track classification faces a challenge with the fixed content--track relationship, caused by consistent instrument type within a track throughout a performance, leading to homogenize feature map, biasing the attention-based classifier to make predictions based on simple track ID cues rather than actual audio content. We introduce track permutation augmentation to overcome this limitation. In training, track IDs are randomly permuted and reassigned across all tracks, with corresponding label adjustments. This approach effectively prevents the model from relying on track IDs for classification, thereby making the learning process more efficient and enhancing performance, while preserves the integrity of the multitrack data.

%% file: sections/3_experiment.tex
\section{Experiments}

\subsection{Implementation Details}
\label{sec:implementation}
We primarily utilize the MJN dataset for our experiments because it is organized by performance and minimizes instrument changes, which simplifies the control of testing conditions. The validation set includes three challenging performances to test model robustness, while the test set features two typical band settings with a relatively balanced label distribution. The remaining 20 performances make up the training set. For MedleyDB, data is split at the song level, with 15\% randomly allocated to both validation and test sets.

Audio are normalized to -0.1dB, converted to mono, resampled to 24000 Hz, and segmented into 5-second clips with 2.5-second overlaps. We use \texttt{MERT-v1-95M}\footnote{\url{https://huggingface.co/m-a-p/MERT-v1-95M}} as our audio encoder with full parameter fine-tuning. The track-wise attention is implemented with a 12-head multihead attention, with layer normalization and large dropout (p=0.8) afterwards. 

Training utilizes the AdamW optimizer \cite{loshchilov2017fixing} with a weight decay of 0.01. Learning rates are set at 1e-5 for both the audio encoder and track-wise attention, and 1e-3 for the linear classifier. Cross entropy loss is used as the objective function. The model undergoes two training epochs with a batch size of 4, supplemented by 4-step gradient accumulation to achieve a total batch size of 16. Validation occurs every quarter epoch, with checkpoint selection based on the Macro F1 score on the validation set. Training is performed on an RTX 3090 GPU (24GB).

\subsection{Metrics}

We utilize accuracy and Macro F1 score as metrics, calculated directly at the frame level. Each metric is first averaged over a 5-second sample and then across the entire test set. Accuracy measures the percentage of correct detections made by the model, while Macro F1 is chosen for its ability to equally weigh the performance of each class---crucial in our datasets, where lead instrument distribution is imbalanced. For instrument classification models, we calculate instrument F1 (\textbf{Inst F1}) and accuracy (\textbf{Inst Acc}) directly. For track classification models, we first compute Macro F1 for track predictions (\textbf{Track F1}), then map these to instrument types using the known track--instrument relationships to compute instrument-level Macro F1 and accuracy, facilitating cross-scheme comparisons.

\subsection{Baseline Models}
To show the effectiveness of our track-wise attention, we implement two variants of our model: \textbf{From mix}, a straightforward implementation using a MERT model with a linear classifier processing only the mixture track; and \textbf{Track avg.}, where the classifier operates on the averaged feature maps from all tracks. Finally, \textbf{Track attn.} is our model incorporating track-wise attention.

Additionally, we compared against two external models. The first is a CRNN model from \cite{adavanne2018multichannel}, designed for sound event detection tasks with multi-channel audio as inputs. The second baseline is an SVM model from \cite{pati2017dataset}, which focuses on segment-level binary classification of guitar solos using mixture audio. We adapt our ``From mix'' model for segment-level classification by implementing average pooling across all frames in the feature map before classification with a linear classifier.

%% file: sections/4_results.tex
\section{Results}

\subsection{Comparison on Classification Module Design}
\input{tables/tb_comparison}

As Table~\ref{tab:comparison} illustrates, the \textit{track avg.} model aid performance with signal from instrument track to achieve better performance than the \textit{from mix} model, but without enough robustness as the performance enhancement varies significantly between easier (+15.85\% instrument F1 on the test set) and challenging cases (+2.56\% on the validation set). Model with track-wise attention more effectively utilizes track information, as evidenced by gains in both validation (+8.08\%) and test (+10.03\%) sets, compared to \textit{track avg.} model. Moreover, switching to track classification further enhances performance, particularly in difficult cases, with a notable +6.24\% increase in instrument F1 on the validation set. Overall, the combination of track-wise attention and track classification gives the strongest performance.\footnote{Recordings in the test set lack multiple instances of the same instrument type within each performance, leading to identical Track F1 and Inst F1 values.}

Additionally, we demonstrate the generalization capability of track classification using the organ---an instrument not present in the training set, with timbre similar to that of the electric guitar. While instrument classification frequently mislabels it as electric guitar (75.68\%), with no correct classifications, track classification achieves 62.42\% accuracy, showcasing enhanced adaptability to unseen instruments.


\subsection{Ablation Studies}
\input{tables/tb_ablation}
Table~\ref{tab:ablation} presents three sets of ablation study results. First, we assess the impact of omitting key components. The removal of track permutation significantly diminishes performance, particularly in challenging scenarios, and fails to achieve results comparable to the model with instrument classification in Table~\ref{tab:comparison}. Similarly, excluding instrument embeddings leads to substantial performance losses. Eliminating track embeddings causes the track classification model to nearly fail at making predictions.

We then explore fine-tuning configurations to justify our choice of full parameter fine-tuning. Freezing the MERT model did not yield meaningful results, while fine-tuning its last transformer layer along with the attention and classifier significantly boosted performance, approaching that of our final model. This indicates that training more parameters in the audio encoder leads to higher performance.

Additionally, we evaluate the model's dependency on the human-produced mixture track by excluding it (\textit{w/o oracle mix}) and replacing it with a pseudo mixture track, created by averaging all single-instrument tracks at the waveform level. This adjustment results in only a very slight performance decrease, suggesting that our model remains functional without the human-produced mixture track, making it applicable in fully automated scenarios.

\subsection{Cross-Dataset Testing}
\input{tables/tb_cross}
Table~\ref{tab:cross}\footnote{This experiment used a smaller batch size (=1), leading to slightly weaker performance on the MJN test set compared to Table~\ref{tab:comparison}.} presents the results for out-of-domain testing. When trained on the MJN dataset, the track classification model not only maintains strong in-domain performance but also excels in out-of-domain tests, showing a +16.45\% improvement in instrument F1 on MedleyDB over the instrument classification model. Training with both datasets yields the best overall performance, highlighting the importance of data quantity and diversity. Moreover, training with MedleyDB and testing on MJN results in a significant performance drop (-32.26\% in instrument F1), but this decline is less pronounced when reversing the training and testing sets (-10.34\%), suggesting MJN is a more effective training set for this task. This observation highlight key strategies for future dataset construction with limited resources: tolerating data imperfections like bleeding sound and less instrument diversity, ensuring a more balanced distribution of lead instruments and more frequent switches of lead instruments.

\subsection{Comparison with Prior Works}

\subsubsection{CRNN}
\input{tables/tb_crnn}
The comparison with CRNN is presented in Table~\ref{tab:crnn}. Overall, the CRNN model performs poorly on our task, regardless of whether the audio features used are mel spectrograms or MERT features, and whether MERT is fine-tuned or not. The performance gap between CRNN and our model is substantial.

\subsubsection{SVM}
\input{tables/tb_svm}
As shown in Table~\ref{tab:svm}, the SVM model also struggles on our dataset, achieving only 26.03\% F1 for guitar. In contrast, our model maintains competitive performance in segment-level classification, demonstrating a significant advantage over the SVM model.

%% file: tables/tb_comparison.tex
\begin{table}[tb]
\centering
\caption{Comparison of classification module designs. Cls. refers to classification schemes.}
\label{tab:comparison}
\resizebox{\columnwidth}{!}{%
\begin{tabular}{cc|ccc|ccc}
\hline
\multicolumn{1}{l}{} & \multicolumn{1}{l|}{} & \multicolumn{3}{c|}{\textbf{Validation set (hard)}} & \multicolumn{3}{c}{\textbf{Test set (balanced)}} \\
\textbf{Model} & \textbf{Cls.} & \textbf{Track F1} & \textbf{Inst F1} & \textbf{Inst Acc} & \textbf{Track F1} & \textbf{Inst F1} & \textbf{Inst Acc} \\ \hline
From mix & Inst. & - & 66.78 & 73.88 & - & 57.17 & 72.69 \\
Track avg. & Inst. & - & 69.34 & 75.57 & - & 73.02 & 83.67 \\
Track attn. & Inst. & - & 77.42 & 83.96 & - & 83.32 & \textbf{91.28} \\
Track attn. & Track & 80.56 & \textbf{83.66} & \textbf{87.79} & 83.76 & \textbf{83.76} & 91.20 \\ \hline
\end{tabular}%
}
\end{table}

%% file: tables/tb_ablation.tex
\begin{table}[tb]
\centering
\caption{Ablation studies}
\label{tab:ablation}
\resizebox{\columnwidth}{!}{%
\begin{tabular}{c|ccc|ccc}
\hline
\multicolumn{1}{l|}{} & \multicolumn{3}{c|}{\textbf{Validation set (hard)}} & \multicolumn{3}{c}{\textbf{Test set (balanced)}} \\
\textbf{Model} & \textbf{Track F1} & \textbf{Inst F1} & \textbf{Inst Acc} & \textbf{Track F1} & \textbf{Inst F1} & \textbf{Inst Acc} \\ \hline
Ours & \textbf{80.56} & \textbf{83.66} & \textbf{87.79} & \textbf{83.76} & \textbf{83.76} & \textbf{91.20} \\ \hline
w/o track perm & 63.71 & 69.55 & 70.20 & 78.44 & 78.44 & 89.59 \\
w/o inst emb & 68.15 & 71.60 & 74.29 & 78.88 & 78.88 & 89.36 \\
w/o track emb & 33.64 & 42.16 & 35.53 & 21.38 & 21.38 & 24.86 \\ \hline
Freeze MERT & 31.13 & 34.98 & 47.71 & 26.41 & 26.41 & 38.30 \\
FT last layer & 56.89 & 60.59 & 74.51 & 43.87 & 43.87 & 71.31 \\ \hline
w/o oracle mix & 80.48 & 82.94 & 86.88 & 83.08 & 83.08 & 90.44 \\ \hline
\end{tabular}%
}
\end{table}

%% file: tables/tb_cross.tex
\begin{table}[tb]
\centering
\caption{Cross-dataset testing results}
\label{tab:cross}
\resizebox{\columnwidth}{!}{%
\begin{tabular}{lcc|cc|cc}
\hline
 & \multicolumn{2}{c|}{\textbf{Training set}} & \multicolumn{2}{c|}{\textbf{Test set (MJN)}} & \multicolumn{2}{c}{\textbf{Test set (MedleyDB)}} \\
\multicolumn{1}{c}{\textbf{Cls}} & \textbf{MJN} & \textbf{MedleyDB} & \textbf{Inst F1} & \textbf{Inst Acc} & \textbf{Inst F1} & \textbf{Inst Acc} \\ \hline
Inst. & Y & \multicolumn{1}{l|}{} & 82.04 & 90.53 & 57.68 & 63.06 \\
Track & Y & \multicolumn{1}{l|}{} & 81.34 & 89.50 & 74.13 & 69.01 \\
Track & \multicolumn{1}{l}{} & Y & 52.30 & 60.16 & \textbf{84.72} & 78.87 \\
Track & Y & Y & \textbf{84.56} & \textbf{92.11} & 84.47 & \textbf{83.25} \\ \hline
\end{tabular}%
}
\end{table}

%% file: tables/tb_crnn.tex
\begin{table}[tb]
\centering
\caption{Comparison with CRNN}
\label{tab:crnn}
\resizebox{\columnwidth}{!}{%
\begin{tabular}{lcc|cc|cc}
\hline
 & \multicolumn{1}{l}{} & \multicolumn{1}{l|}{} & \multicolumn{2}{c|}{\textbf{Validation set (hard)}} & \multicolumn{2}{c}{\textbf{Test set (balanced)}} \\
\multicolumn{1}{c}{\textbf{Model}} & \textbf{Input} & \textbf{FT MERT} & \textbf{Inst F1} & \textbf{Inst Acc} & \textbf{Inst F1} & \textbf{Inst Acc} \\ \hline
Ours & MERT feat. & Y & \textbf{77.42} & \textbf{83.96} & \textbf{83.32} & \textbf{91.28} \\
CRNN & Mel spec. & NA & 37.64 & 46.36 & 17.15 & 24.93 \\
CRNN & MERT feat. & N & 57.62 & 60.16 & 23.20 & 26.48 \\
CRNN & MERT feat. & Y & 57.62 & 60.16 & 23.20 & 26.48 \\
CRNN + attn. & MERT feat. & Y & 57.62 & 60.16 & 26.50 & 31.40 \\ \hline
\end{tabular}%
}
\end{table}

%% file: tables/tb_svm.tex
\begin{table}[tb]
\centering
\caption{Comparison with SVM on segment-level guitar solo detection}
\label{tab:svm}
\resizebox{\columnwidth}{!}{%
\begin{tabular}{c|ccc|ccc}
\hline
\multicolumn{1}{l|}{} & \multicolumn{3}{c|}{\textbf{Validation set (hard)}} & \multicolumn{3}{c}{\textbf{Test set (balanced)}} \\
\textbf{Model} & \textbf{Acc} & \textbf{Guitar F1} & \textbf{Macro F1} & \textbf{Acc} & \textbf{Guitar F1} & \textbf{Macro F1} \\ \hline
Ours & \textbf{93.02} & \textbf{82.17} & \textbf{88.91} & \textbf{92.20} & \textbf{87.50} & \textbf{90.91} \\
SVM & 75.50 & 26.03 & 55.67 & 56.44 & 20.80 & 45.38 \\ \hline
\end{tabular}%
}
\end{table}

%% file: sections/5_conclusion.tex
\section{Conclusion}

In this paper, we introduced the task of lead instrument detection in multitrack music and developed two annotated datasets specifically for this purpose. Our proposed model, which incorporates an SSL audio encoder, instrument and track embeddings, track-wise attention, track classification, and track permutation augmentation, effectively addresses this task and is capable of generalizing to unseen instruments. We established robust baselines on these datasets and demonstrated the superiority of our approach through comparative studies with CRNN and SVM models. This exploration of multitrack audio content analysis provides valuable insights for future research on similar tasks.